\documentclass{PoS}

\title{Magnetized, Relativistic Jets}

\ShortTitle{Magnetized, Relativistic Jets}

\author{\speaker{Manel Perucho}%
        \thanks{MP and J-MM acknowledge financial support
from the Spanish Ministerio de Econom\'{\i}a y Competitividad (grants
AYA2013-40979-P, and AYA2013-48226-C3-2-P) and from the local
{Valencian Government} (Generalitat Valenciana, grant
Prometeo-II/2014/069). JLG and AF acknowledge support from the Spanish
Ministerio de Econom\'{\i}a y Competitividad grants
AYA2013-40825-P and AYA2016-80889-P. }\\
       Departament d'Astronomia i Astrof\'{\i}sica. Universitat de Val\`encia\\
       E-mail: \email{manel.perucho@uv.es}}

\author{Jos\'e M. Mart\'{\i}\\
        Departament d'Astronomia i Astrof\'{\i}sica. Universitat de Val\`encia\\
        E-mail: \email{jose-maria.marti@uv.es}}
        
\author{Jos\'e L. G\'omez\\
        Instituto de Astrof\'{\i}sica de Andaluc\'{\i}a\\
        E-mail: \email{jlgomez@iaa.es}}        
        
\author{Antonio Fuentes\\
        Instituto de Astrof\'{\i}sica de Andaluc\'{\i}a\\
        E-mail: \email{afuentes@iaa.es}}

\abstract{Extragalactic relativistic jets are composed by charged particles and magnetic fields, as inferred from the synchrotron emission that we receive from them. The Larmor radii of the particles propagating along the magnetic field are much smaller than the scales of the problem, providing the necessary coherence to the system to treat is as a flow. We can thus study them using relativistic magnetohydrodynamics. As a first step, we have studied the structure of steady-state configurations of jets by using numerical simulations. We have used a helical field configuration and have changed different relevant parameters that control the way in which the energy flux is distributed in jets (namely, the proportion of the energy flux carried by internal, kinetic or magnetic energy). Our results show significant differences among the different kinds of jets. Finally, we also report on results based on synthetic maps of our simulated jets.}

\FullConference{XII Multifrequency Behaviour of High Energy Cosmic Sources Workshop\\
         12-17 June\\
         Palermo, Italy}

\begin{document}

\section{Introduction}
Extragalactic, relativistic jets are associated to radio-emitting active galactic nuclei. According to state-of-art simulations (e.g., \cite{MB09,TN11,PO13}), relativistic jets originate from the extraction of rotational energy from the supermassive black hole (SMBH) by magnetic field lines, i.e., the Blandford-Znajek model \cite{BZ77}. A continuous supply of plasma to the SMBH ensures a continuous presence of magnetic fields that cross the ergosphere of the SMBH. The twisted magnetic lines generate an outwards pressure that pushes particles out along the rotation axis. Recent observational results show that the magnetic field close the the galactic nucleus is of the expected order to guarantee the formation of jets within the model \cite{ZC14,BA16}

An alternative (or complementary) way to produce jets and outflows is given by the extraction of particles from the accretion disk along poloidal field lines \cite{BP82}. This mechanism tends to produce slower and denser outflows, as a consequence of the proton loading from the disk. It is out of the scope of this paper to discuss the possible two-fluid structure of jets, which implies both formation mechanisms acting at the same time and producing a fast and collimated inner spine surrounded by a slower and denser wind coming from the disk. We focus on the inner spine of superfast magnetosonic jets. It is also interesting to mention that there are other formation models that imply completely different jet structures to those that we deal with here, considering that an electromagnetic drift is responsible for the launching of all astrophysical jets (e.g., \cite{KGK04}). 

The presence and structure of magnetic fields in jets at parsec-scales has also been outlined in many observational papers, suggesting that the field must have a structured helical morphology, probably modulated by a turbulent component (e.g., \cite{GA15,GL16}). 

We present here a summary of our work on the structure of magnetized relativistic jets using numerical simulations. The numerical simulations have been run using the code presented in \cite{ma15a,ma15b}. In particular we focus on the effect of the dominating energy channel on the jet structure. The energy flux in jets is distributed among internal, kinetic and magnetic energy fluxes. Therefore, we vary the role of each of them and analyse the implications for jet structure. From our results, we provide a method to give upper limits on the jet magnetosonic Mach number.

\section{Jet structure in a steady-state situation}
We focus here on the role of magnetic fields on the transversal structure of jets. The RMHD equations in (orthonormal) cylindrical coordinates (see, e.g., the Appendix~A in \cite{LA05} and \cite{ma15b}) reduce to a single ordinary differential equation for the transversal equilibrium of the jet, in a steady state situation and considering slab symmetry along the z axis,
\begin{equation}
\label{eq:teq}
  \frac{d p^*}{d r} = \frac{\rho h^* W^2 (v^\phi)^2
  - (b^\phi)^2}{r}.
\end{equation}
where $p^*$ and $h^*$ stand for the total pressure and the
specific enthalpy including the contribution of the magnetic field
\begin{equation}
\label{eq:p*}
  p^* = p + \frac{b^2}{2}
\end{equation}
\begin{equation}
\label{eq:h*}
  h^* = 1 + \varepsilon + p/\rho + b^2/\rho,
\end{equation}
where $p$ is the gas pressure, $\rho$ its density and $\varepsilon$ its specific
internal energy, $b^\mu$ ($\mu = t,r,\phi,z$) are the components of the magnetic field 4-vector in the fluid rest frame, $v^i$ ($i = r, \phi,
z$) are the components of the fluid 3-velocity in the laboratory
frame, and $W$ is the flow Lorentz factor. Throughout this text, the speed of light is taken to be $c=1$. We can relate the magnetic field components in the reference frame to its 3-vector components in the lab frame, $B^i$ using the following relations:
\begin{eqnarray}
\label{b0}
  b^0 & = & W B^i v_i \ , \\
  \label{bi}
  b^i & = & \frac{B^i}{W} + b^0 v^i.
\end{eqnarray}
The square of the modulus of the
magnetic field is thus 
\begin{equation}
  b^2 = \frac{{B}^2}{W^2} + (B^i v_i)^2 
\end{equation}
with $B^2 = B^iB_i$ (summation over repeated indexes is assumed).

\section{Results}

  The results of the simulations run can be found in \cite{mpg16}. However, we focus here on a careful discussion of jet structure based on Equation (\ref{eq:teq}). This equation establishes the transversal equilibrium
between the total pressure gradient and the centrifugal force (first
term on the right hand side), that tends to produce a positive gradient of the
radial total pressure profile, and the magnetic tension (second term
on the r.h.s), which favours to increase the total pressure
towards the axis. 
  
  Real, extragalactic jets are observed to undergo flaring events which imply strong dynamical changes of its internal structure. However, we do not observe jets strongly expanding and contracting between epochs. Therefore, we assume here that the jet is in equilibrium with the ambient medium at each position. This justifies the following discussion of jet structure in which we use Equation (\ref{eq:teq}) in the context of interferometric radio observations.

\subsection{Jet structure}

If we remove the centrifugal term on the r.h.s of Equation (\ref{eq:teq}), and considering the jet to be in pressure equilibrium with the ambient medium, the transversal structure of the jet is determined by the total pressure gradient and the magnetic tension. This means that the strongest the toroidal field is, the steeper the negative pressure gradient in the radial direction, and the larger the pressure on the jet axis to prevent the jet from being pinched by the toroidal field.  

 Therefore, if the toroidal magnetic field is small enough the gas pressure profile will be flatter inside
the jet, whereas if the magnetic tension is more relevant, the gas can drop several orders of magnitude across the jet, producing an equilibrium model with a central spine with high pressure \cite{ma15b}. Figure~1 shows this effect for two jets, with weak and strong toroidal fields.

\begin{figure}
\includegraphics[width=\textwidth]{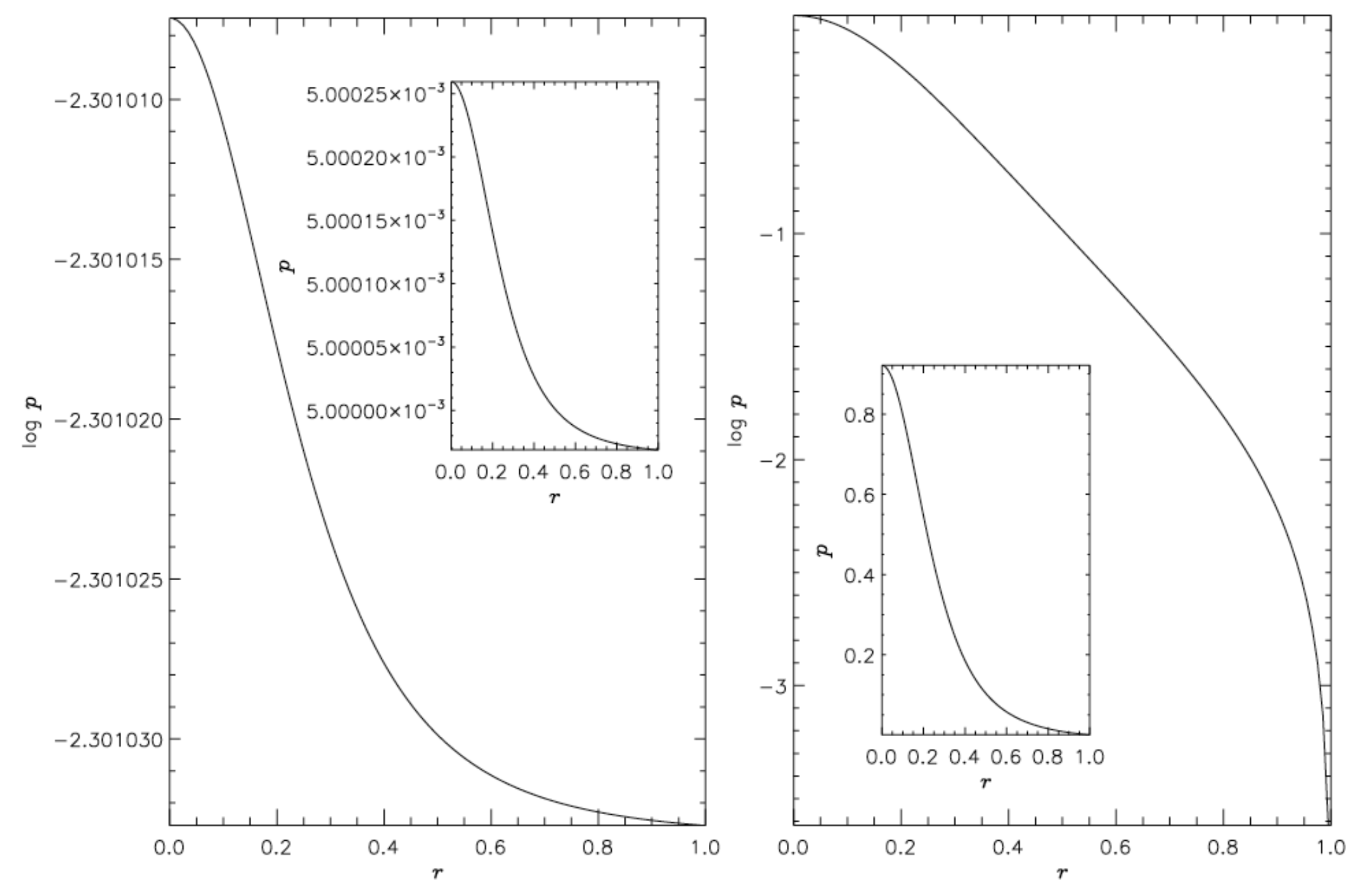}
\caption{Pressure profile for two cases of jets in equilibrium, with weak (left panel) and strong (right panel) toroidal fields \cite{ma15b}.}
\label{fig1}
\end{figure}

The numerical simulations presented in \cite{mpg16} show this effect, which can be attached to high pressure within the jet spine. This spine pressure can be both given by large values of gas pressure, by a strong poloidal magnetic field,\footnote{Symmetry forces the toroidal field to vanish on the axis.} or a combination of both.  

As stated above, the centrifugal term in Equation (\ref{eq:teq}) triggers a positive gradient, and, depending on the profile of $v^\phi$ and $b^\phi$, the pressure profile will have an increasing, decreasing or variable trend that may show deep drops in total pressure at certain jet radii (see \cite{ma15b}). The competition between both terms can give rise to deep minima in the gas pressure profile at the locations in which there is a transition from regions dominated by tension to regions dominated by rotation across the jet (see Fig.~2).
  
\begin{figure}
\includegraphics[width=\textwidth]{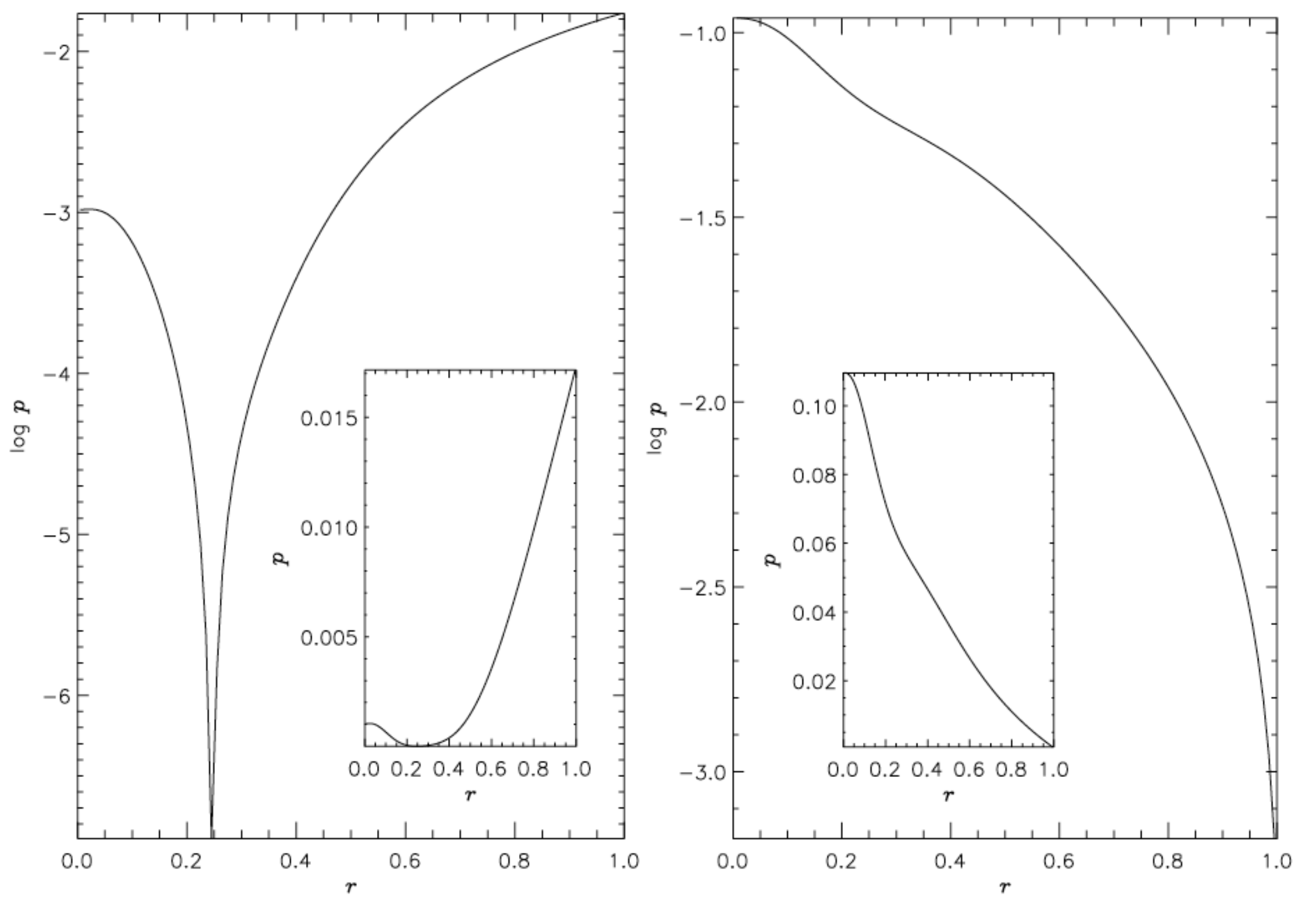}
\caption{Pressure profile for two cases of jets in equilibrium, with weak (left panel) and stronger (right panel) toroidal fields, including differential toroidal velocity \cite{ma15b}.}
\label{fig2}
\end{figure}

\subsection{Departure from equilibrium}

  If the (superfast magnetosonic) jet is overpressured (underpressured) with respect to its environment, it expands (contracts) and recollimates (re-expands), forming subsequent reconfinement shocks. We studied the internal structure of a number of jets with the same Lorentz factor ($W=3.5$) and adiabatic exponent ($\gamma=4/3$) distributed across the parameter space shown in Fig.~3.

 \begin{figure}
 \begin{center}
\includegraphics[width=\textwidth]{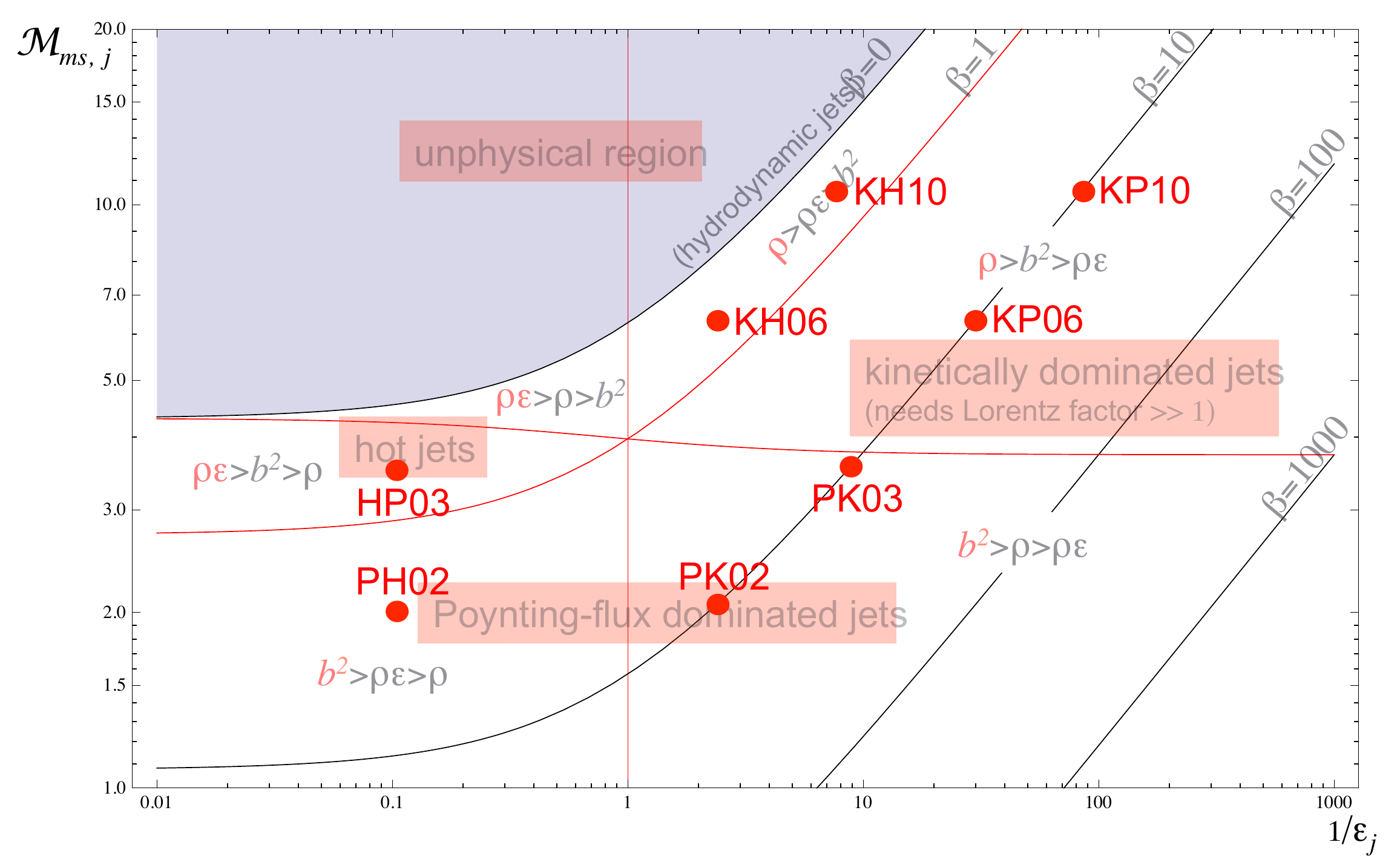}
\caption{Distribution of some of the studied jet models across the (inverse of the) specific internal energy vs. (relativistic) magnetosonic Mach number plane \cite{mpg16}.}
 \end{center}
\label{fig3}
\end{figure}

  Our simulations show that the jet structure is completely dependent on the energetic distribution of the jet within the different energy channels.  It is important to stress here that we have not considered a set of jets with the same energy flux, but that they have been chosen by fixing the jet Lorentz factor and rest-mass density and changing the relative proportion of internal and/or magnetic energy. Taking into account that the separation between reconfinement shocks depends on the magnetosonic Mach number of the flow, the general trends that we find are the following: 1) if the jet energy flux is dominated by the internal energy, the flow accelerates along expansion regions due to the Bernoulli theorem, and the recollimation shocks appear closer to each other (see Fig.~4); 2) if the flux is dominated by the magnetic energy, the portion of the flux into internal energy still has an influence on the distance between recollimation shocks (the hotter the jet, the closer the shocks appear), and the tension generated by the toroidal field causes a strong increase of the pressure close to the jet axis (see Fig.~5), and 3) if the flux is dominated by the kinetic energy, the distance between the shocks increases, and basically no acceleration is observed along expansion regions (see Fig.~6). 
       
 \begin{figure}
 \begin{center}
\includegraphics[width=\textwidth]{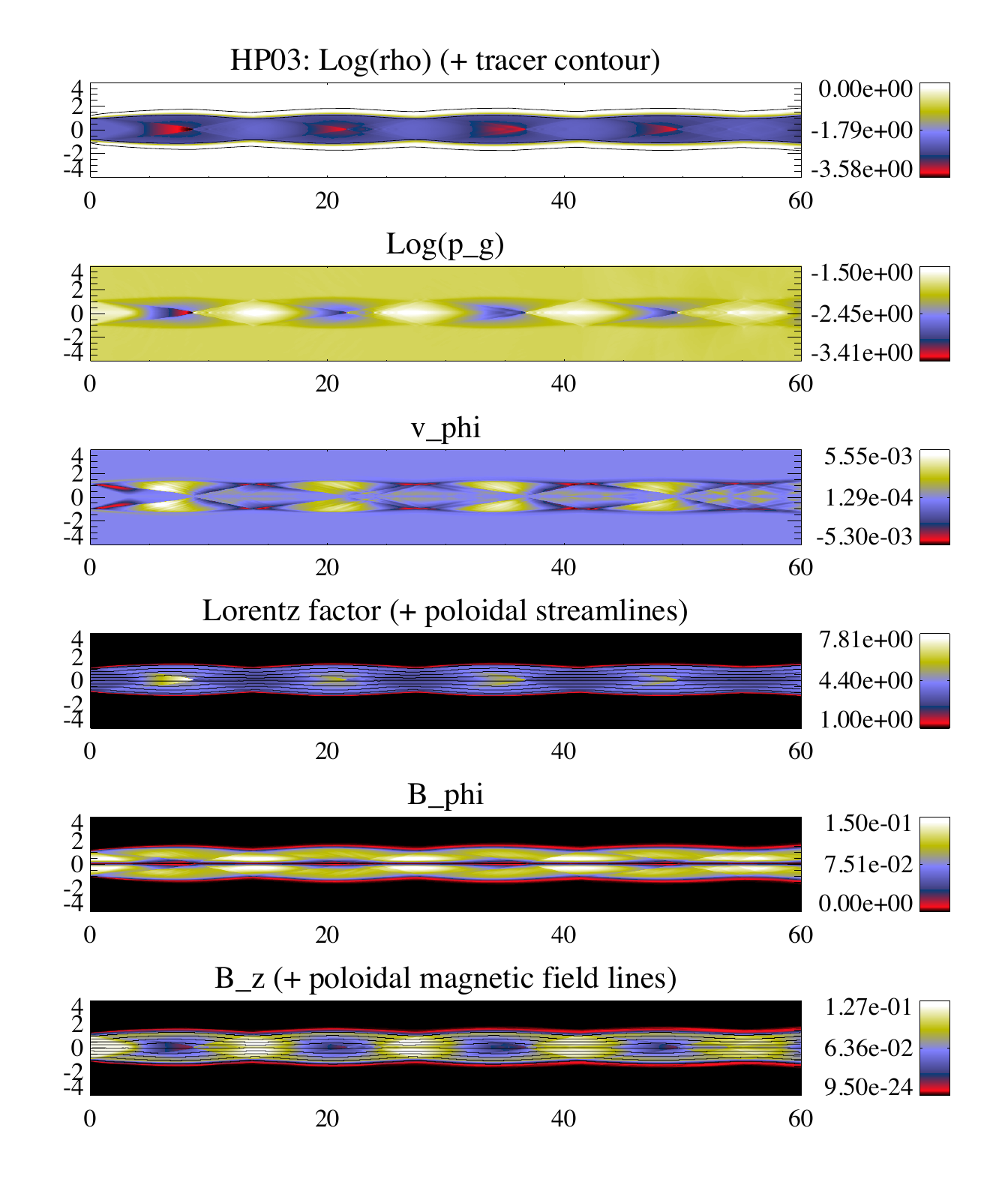}
\caption{Steady state jet structure for a hot jet flow, HP03 in Fig.~3 \cite{mpg16}.}
 \end{center}
\label{fig4}
\end{figure}
 
 \begin{figure}
  \begin{center}
\includegraphics[width=\textwidth]{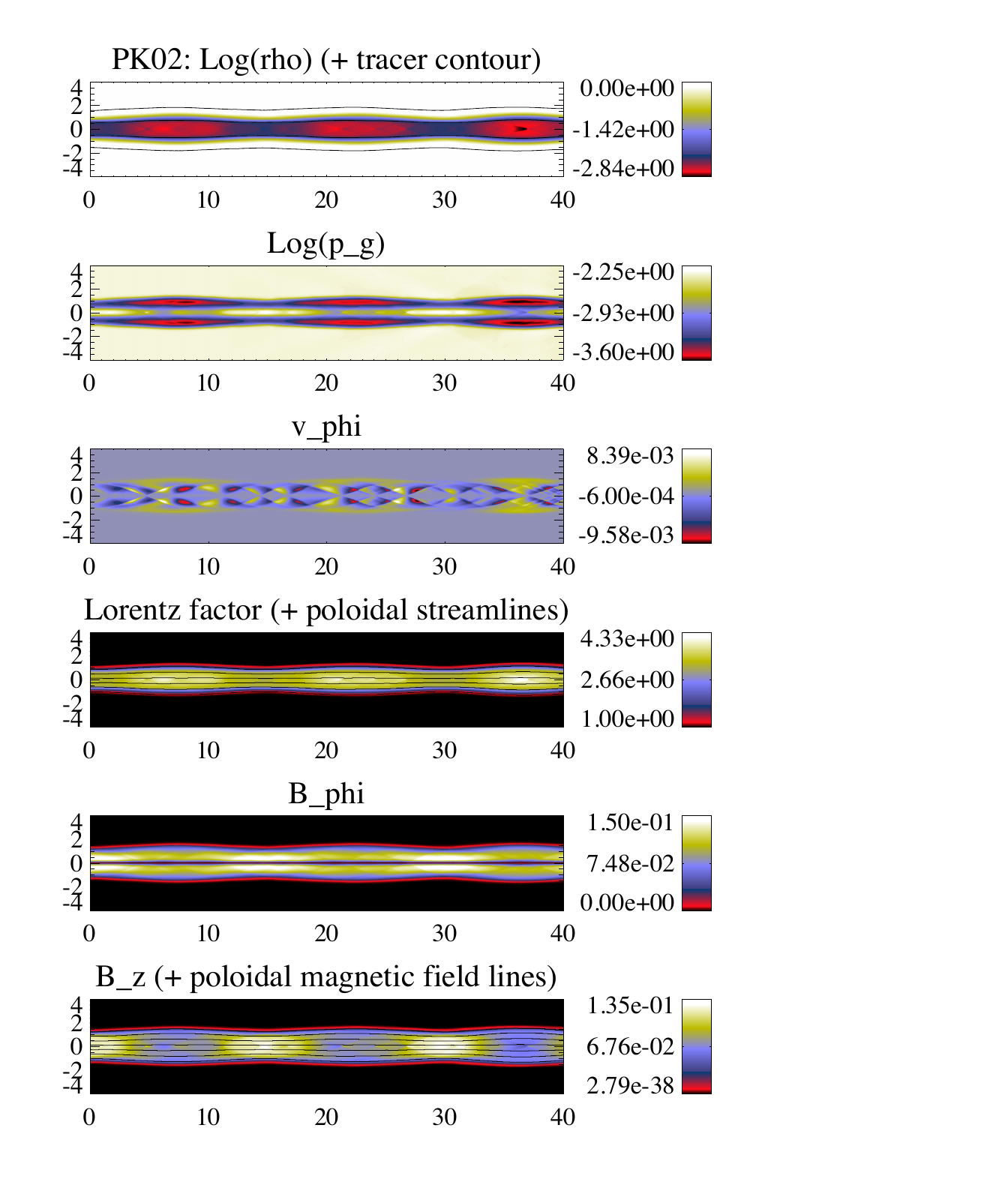}
\caption{Steady state jet structure for a magnetically dominated jet flow, PK02 in Fig.~3  \cite{mpg16}.}
   \end{center}
\label{fig5}
\end{figure}
 
 \begin{figure}
  \begin{center}
 \includegraphics[width=\textwidth]{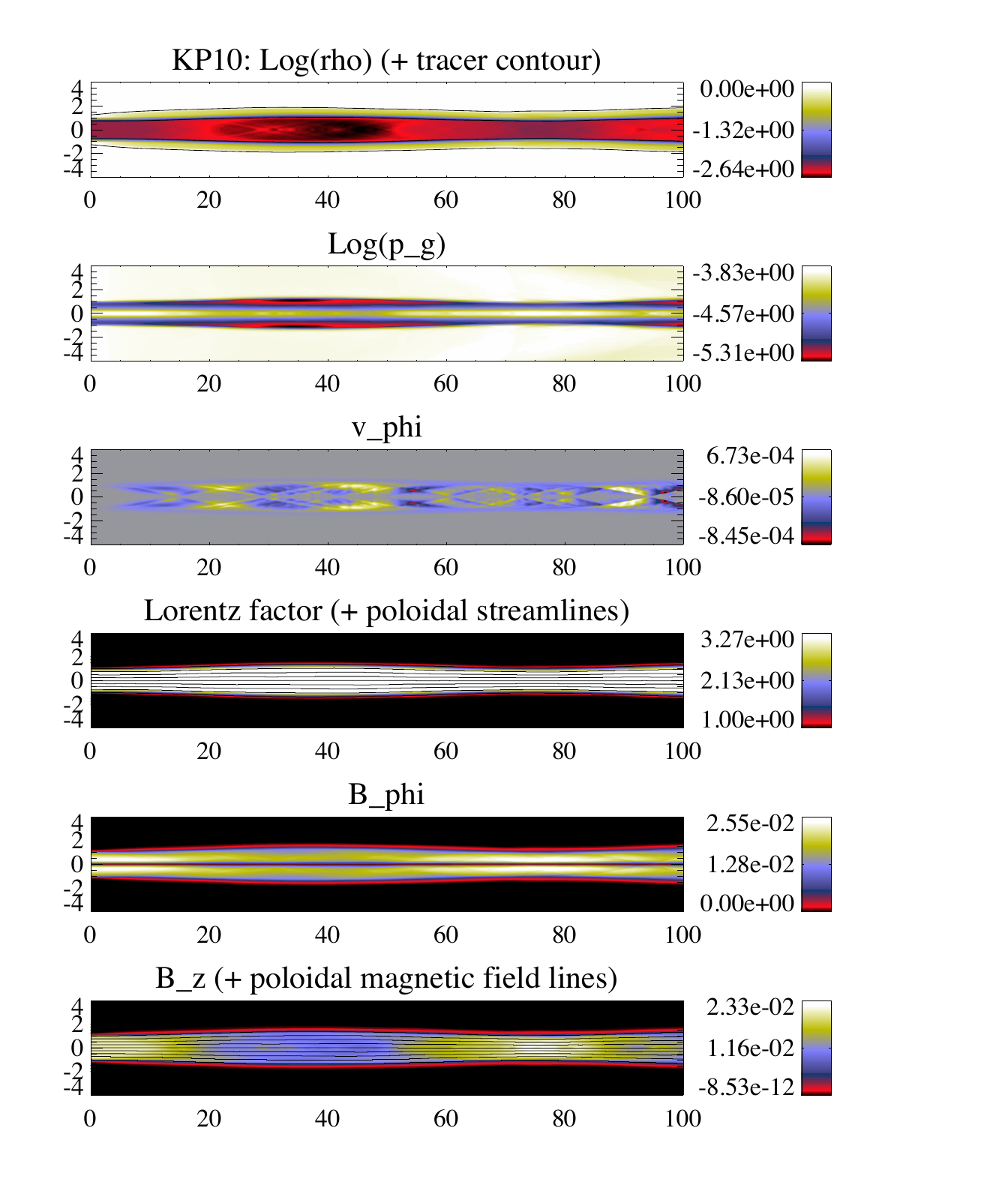}
\caption{Steady state jet structure for a kinetically dominated jet flow, KP10 in Fig.~3 \cite{mpg16}.}
  \end{center}
\label{fig6}
\end{figure}
  
Interestingly, the numerical simulations also showed that strongly magnetized jets in which the toroidal and poloidal field components rapidly drop at the jet boundary are prone to the development of current-driven instabilities that disrupt the jet \cite{ma16}. This points towards small values of toroidal field intensity or thick shear layers in order to keep jets collimated through long distances. 
  
  One of the consequences of jet expansion and contraction is the enhancement of flow rotation. The Lorentz force acting on the relativistic magnetized fluid is 
\begin{equation}
{\bf F}_L = {\bf J} \times {\bf B} + \rho_e {\bf E},
\end{equation}
where ${\bf J} = \nabla \times {\bf B}$ and $\rho_e = \nabla \cdot
{\bf E}$ are the current and electric charge densities, and ${\bf E} = - {\bf v} \times {\bf B}$ is
the electric field (in the ideal MHD approximation). In cylindrical
coordinates $(r, \phi,z)$, and for an axisymmetric flow, the azimuthal
component of the Lorentz force is
\begin{equation}
F_L^\phi  =  B^z \displaystyle{\frac{\partial B^\phi}{\partial z}}  + \frac{B^r}{r}
\frac{\partial (rB^\phi)}{\partial r} + \rho_e (v^r B^z + v^z B^r).
\end{equation}
 
 Therefore, azimuthal velocity can be triggered/enhanced by the changes in the jet width (see Figs.~4 and 5). The azimuthal velocity can, in turn, enhance brightness asymmetries in the jet, on top of those given by a helical magnetic field, e.g., \cite{AG00,LY05,CB11} and Fuentes et al. (in preparation).
 
  The distance between the shocks depends on the jet and ambient medium properties \cite{DM88,FA91}. In particular, it depends on the opening angle of the jet, which is approximately the inverse of the magnetosonic Mach number (for small enough opening angles) if the jet expands freely. If the ambient medium has a fairly constant pressure between shocks, as it is the case in our simulations, the jet expansion and collimation are symmetric, so one can relate the distance between shocks and jet width to the magnetosonic Mach number ($M_{ms}$) of the flow, with $\tan (\alpha) = 1/M_{ms}$. This is certainly the reason behind the results summarised above regarding the distance between shocks in different kinds of jets.  
  
  Our simulations show that there is indeed a clear correlation between the shock positions and $M_{ms}$ of the flow \cite{mpg16}. Therefore, the argumentation can be applied to real extragalactic jets that show stationary components that can be attached to reconfinement shocks. If the jet is transversally resolved by the VLBI observation, the largest jet radius at its maximum expansion divided by the distance between components can give an accurate idea of $M_{ms}$. If the jet is not resolved, the size of the fitted stationary components in the radio map provide us with a minimum transversal jet size, which in turn gives a lower limit to the opening angle and thus, an upper limit to $M_{ms}$.    
 
 \subsection{Emission calculations}

 Work in preparation includes the computation of the emission derived from our RMHD simulations (Fuentes et al., in preparation) using radiative transfer equations in the same way as it is done in \cite{GM95,GM97}. The emission simulations show different effects of the field on the observed emission, namely: 
 \begin{itemize}
 \item A bright spine is generated by a strong toroidal component, as expected from the pressure distribution profile generated (see Fig.~1).
\item The helical magnetic field induces a top/down asymmetry, which reverts at a viewing angle of $\simeq 20^\circ$ for the used pith angle of $45^\circ$ and the common Lorentz factor of 3.5  in our simulation.
\item Small viewing angles show bimodal distributions of EVPAs (aligned or perpendicular to the jet). 
\item Small variations of the EVPAs ($\simeq 15^\circ$) are observed at recollimation shocks.
\item Kinetically dominated flows show stronger relative brightness at recollimation shocks.
 \end{itemize}
 
 \section{Discussion}
 
 Jets are probably formed as Poynting flux dominated, sub-slow magnetosonic flows close to their central engines \cite{KO12}. They must be thermodynamically relativistic too, because of the energy budget and the relatively small amount of particles involved if the jet is formed as a Blandford-Znajek jet. 
 
 At parsec-scales, they show superluminal motions, which implies acceleration and conversion of internal and/or magnetic energy into kinetic energy, and they also show well-collimated structure, with opening angles $< 1$~deg. The collimating mechanism is under debate because, although the tension provided by the toroidal field can do this job, it is also prone to trigger current-driven instabilities in the flow (e.g., \cite{SP97}). The disruptive effect can be relaxed by a thick shear layer/wind with an axial field component that prevents the jet from suffering large amplitude oscillations (also because of the magnetic tension). The ambient medium pressure can also confine the jet, but even if none of both options are valid, the jet's own relativistic nature can confine it, as its opening angle is $\propto 1/W$.
 
 Regarding the acceleration, it has been shown that a cold, magnetised jet can only invest magnetic energy into kinetic energy if its expansion is not homogeneous across the jet surface, i.e., differential expansion and paraboloidal jet shapes would be required \cite{KO12,VK04}. However, if the jet is thermodynamically relativistic, acceleration can still take place in conical jets with constant opening angles across their cross-section. In other words, a cold, conical jet could not accelerate, but a hot, conical jet could accelerate. Although in \cite{VK04} it is claimed that thermal acceleration occurs on small scales, different numerical simulations show that it can extend along many jet radii (e.g., \cite{PM07}). 
 
 Once the jet is accelerated and collimated, its internal structure appears to be completely determined by its pressure ratio with the ambient medium, its Lorentz factor, and its magnetisation-field structure. It is in this region where we have focused our work. We have shown how different distributions of the jet energy flux in different channels can influence the periodicity of recollimation shocks triggered by pressure differences between the jet and the ambient medium, or the stability properties of the flow. Although we have studied a number of models with a fixed helical field structure, i.e., with a common pitch angle, it was shown in \cite{ma15a}, using Eq.~\ref{eq:teq}, that it is very difficult to obtain (non-force-free) equilibrium models for larger magnetisations and pitch angles, in agreement with the expectation of toroidal fields being related to more unstable jet configurations. This has to be overcome by the presence of a shear layer, a decrease of the pitch angle (i.e., an increase of the poloidal field intensity), or by an increase of the jet Lorentz factor (see also \cite{tb16}).
 
Connecting to the largest scales, a number of observations show that the magnetic field structure of FRI jets is probably dominated by toroidal fields, unlike that of FRII jets, which shows a predominance of the poloidal field \cite{GL11,SS88}. Numerical simulations of magnetised, classical and relativistic jets at kiloparsec scales seem to explain this effect by a stretching of the magnetic lines in fast jets \cite{HE11,HK14}.    

Future work in this line of research should include a study of the interaction between travelling perturbations and the steady-state jet structure, focusing on shock-shock interaction (see, e.g., \cite{GM97,FP16} for the RHD counterpart), stability analysis of different jet structures, and kiloparsec scale simulations of magnetised jets in realistic environments (see, e.g., \cite{HK14}).

\bigskip
\bigskip
\noindent {\bf DISCUSSION}

\bigskip
\noindent {\bf WOLFGANG KUNDT:} Manel, in your talk you kindly quoted my 2004 paper with Gopal-Krishna. In it, we try to model all the observed astrophysical jets, with the understanding that 'nature' must be able to blow them in a stable way --hundreds of them-- without a built-in DNA (like in Biology), and we restrict all considerations to 'naked jets' propagating through vacuum. In your talk, you considered a vastly larger set of possibly manmade jets.

\bigskip
\noindent {\bf MANEL PERUCHO:} I referred to your paper as an alternative way to explain the nature of jets to what we typically consider, namely, the Blandford-Znajek model. Within this model, and some basic considerations, one can expect sub-parsec and parsec-scale jets to develop helical magnetic fields. The jet evolution along a given distance can also produce differences in the jet composition or energy distribution. The energy 'channels' of a jet are presumably internal, kinetic and magnetic energy. We have just studied different configurations of steady-state jets threaded by helical fields, in terms of the energy distribution in them. Therefore, even if all jets are formed in the same way, as follows from your model, they can be completely different at a distance from the forming region, owing to their different evolution and resulting composition.

%\bigskip
%\noindent {\bf JAMES WHITE's Comment:} I would like to remind that..........

\end{document}